\documentclass{aa}  

\usepackage{graphicx}
\usepackage{txfonts}
\usepackage{booktabs}
\usepackage{natbib}
\usepackage[colorlinks=true,citecolor=blue]{hyperref}
\usepackage[normalem]{ulem}
\usepackage{siunitx}

\usepackage[utf8]{inputenc}
\usepackage[usenames, dvipsnames]{color}
\usepackage{amsmath}
\usepackage[english]{babel}
\usepackage{float}
\usepackage{url}
\usepackage{longtable}
\usepackage{fancyhdr}
\usepackage[version=4]{mhchem}
\usepackage{gensymb}

\def\Msun{\hbox{M$_{\odot}$}}             
\def\Rsun{\hbox{R$_{\odot}$}}
\def\Mjup{\hbox{M$_{\rm Jup}$}}

\def\Mnep{\hbox{M$_{\rm Nep}$}}

\def\hd20{HD~209458\,b}
\def\gj34{GJ~3470\,b}
\def\hat11{HAT-P-11\,b}

\def\halpha{H$\alpha$ }

\newcommand{\Ppone}{\ensuremath{ 0.5568143^{+3.18e-06}_{-3.26e-06}}}
\newcommand{\tzeropone}{\ensuremath{2458386.17146^{+0.0006}_{-0.0006}}}

\newcommand{\Kp}{\ensuremath{25736.67^{+469.28}_{-422.05}}}

\newcommand{\ecc}{\ensuremath{0.017^{+0.009}_{-0.010}}}
\newcommand{\omegab}{\ensuremath{93.11^{+17.76}_{-16.42}}}

\newcommand{\massMj}{\ensuremath{61.6^{+4.0}_{-4.0}}}
\newcommand{\semimajorau}{\ensuremath{0.0098^{+0.0007}_{-0.0008}}}

\newcommand{\radiusrj}{\ensuremath{0.91^{+0.07}_{-0.07}}}
\newcommand{\rhogcm}{\ensuremath{108.5^{+38.4}_{-27.3}}}
\newcommand{\gplanet}{\ensuremath{2095.29^{+1338.65}_{-755.03}}}

\newcommand{\tstar}{TOI-263\xspace}
\newcommand{\tplanet}{TOI-263\,b\xspace}

\newcommand{\teff}{\ensuremath{T_\mathrm{eff}}\xspace}

\begin{document} 

   \title{ESPRESSO Mass determination of TOI-263b: An extreme inhabitant of the brown dwarf desert}
   \titlerunning{Mass determination of TOI-263b}

 \author{E. Palle\inst{\ref{iiac},\ref{iull}}
   \and R. Luque\inst{\ref{iiac},\ref{iull}}
   \and M.~R.~Zapatero~Osorio\inst{\ref{inst:cab}}
   \and H. Parviainen\inst{\ref{iiac},\ref{iull}}
    \and M.~Ikoma\inst{\ref{inst:hongo}}
    \and H.~M.~Tabernero\inst{\ref{inst:porto}}
   \and M.~Zechmeister\inst{\ref{inst:gott}}
    \and A.J.~Mustill\inst{\ref{inst:lund}}
 \and V.S.J. Bejar\inst{\ref{iiac},\ref{iull}}
    \and N.~Narita\inst{\ref{iiac},\ref{inst:komaba},\ref{inst:jst},\ref{inst:abc}}
   \and F. Murgas\inst{\ref{iiac},\ref{iull}}
 }

  \institute{
  	     \label{iiac} Instituto de Astrof\'isica de Canarias (IAC), E-38200 La Laguna, Tenerife, Spain
  	\and 
  	    \label{iull} Deptartamento de Astrof\'isica, Universidad de La Laguna (ULL), E-38206 La Laguna, Tenerife, Spain
	\and 
        \label{inst:cab} Centro de Astrobiolog\'ia (CSIC-INTA), Carretera de Ajalvir km 4, E-28850 Torrej\'on de Ardoz, Madrid, Spain
    \and
        \label{inst:hongo} Department of Earth and Planetary Science, The University of Tokyo, 7-3-1 Hongo, Bunkyo, Tokyo 113-0033, Japan
    \and
        \label{inst:porto} Instituto de Astrof{\'i}sica e Ci{\^e}ncias do Espa\c{c}o, Universidade do Porto, CAUP, Rua das Estrelas, 4150-762 Porto, Portugal        
    \and 
        \label{inst:gott} Institut f\"ur Astrophysik, Georg-August-Universit\"at, Friedrich- Hund-Platz 1, 37077 Göttingen, Germany
    \and 
        \label{inst:lund} Lund Observatory, Department of Astronomy \& Theoretical Physics, Lund University, Box 43, SE-221 00 Lund, Sweden        
    \and
        \label{inst:komaba}Komaba Institute for Science, The University of Tokyo, 3-8-1 Komaba, Meguro, Tokyo 153-8902, Japan
    \and
        \label{inst:jst}JST, PRESTO, 3-8-1 Komaba, Meguro, Tokyo 153-8902, Japan
    \and
        \label{inst:abc}Astrobiology Center, 2-21-1 Osawa, Mitaka, Tokyo 181-8588, Japan
  }

  \date{Received dd February 2020 / Accepted dd Month 2020}

  \abstract
   {The TESS mission has reported a wealth of new planetary systems around bright and nearby stars amenable for detailed characterization of the planet properties and their atmospheres. However, not all interesting TESS planets orbit around bright host stars. \tplanet is a validated ultra-short period substellar object in a 0.56-day orbit around a faint ($V=18.97$) M3.5 dwarf star. 
   The substellar nature of \tplanet was explored using multi-color photometry, which determined a true radius of $0.87 \pm 0.21 R_{J}$, establishing \tplanet's nature ranging from an inflated Neptune to a brown dwarf. 
   The orbital period-radius parameter space occupied by \tplanet is quite unique, which prompted a further characterization of its true nature. Here, we report radial velocity measurements of \tstar obtained with 3 VLT units and the ESPRESSO spectrograph to retrieve the mass of \tplanet. We find that \tplanet is a brown dwarf with a mass of  $61.6 \pm 4.0\,\Mjup$. Additionally, the orbital period of the brown dwarf is found to be synchronized with the rotation period of the host star, and the system is found to be relatively active, possibly revealing a star--brown dwarf interaction. All these findings suggest that the system's formation history might be explained via disc fragmentation and later migration to close-in orbits. If the system is found to be unstable, \tstar is an excellent target to test the migration mechanisms before the brown dwarf becomes “engulfed” by its parent star.
   }

   \keywords{planetary systems -- planets and satellites: individual: TOI-263b  --  planets and satellites: atmospheres -- methods: radial velocity -- techniques:  spectroscopic--  stars: low-mass}
   \maketitle

\section{Introduction}
\label{sec:introduction}

The {\it TESS} mission \citep{Ricker2015} is dedicated to the search for transiting extrasolar planets around the brightest and closest stars in the sky. It was launched on April 18th 2018, and has recently completed  its initial survey of (almost) the entire sky with 26 pointings over 2 years. Many of the new exciting planets discovered by TESS are excellent targets for precise radial velocity (RV) follow-up to determine accurate masses and bulk properties, and later explore the characterization of their atmospheres \citep{Huang2018,Kostov2019, Vanderburg2019, Winters2019, Luque2019}. 

However, despite the thousands of planets discovered to date, many questions remain about their dominant formation mechanism(s) and the underlying statistical properties of the different exoplanet populations \citep{Lissauer2011, Howard2012, Winn2015}. Thus, new objects displaying extreme properties may deliver critical knowledge to understand the formation and evolution of planetary systems. One of those extreme systems is \tplanet (Gaia DR2 5119203027983398656).

\citet{Parviainen2020} confirmed \tstar to be an M3.5 star, with an effective temperature of $\teff = 3250 \pm 140$\,K, a mass of $M = 0.4 \pm 0.1 \Msun$, and a radius of $R=0.405 \pm 0.077 \Rsun$. Further, using ground based multi-color photometry from MuSCAT2 at the TCS telescope \citep{Narita2019}, they excluded contamination from unresolved sources with a significant colour difference from TOI-263, and  validated \tplanet to be a substellar object with a true radius value of $0.87 \pm 0.21 R_{J}$.

Only a few planets and brown dwarfs in short period orbits are known and have their physical parameters measured \citep[][and reference therein]{Deleuil2008, Moutou2013, Csizmadia2015, Persson2019}, and any additional measurements are very valuable for constraining the mass--radius relationship for brown dwarfs and testing substellar evolutionary models  \citep{1997ApJ...491..856B,2000ApJ...542..464C,2008ApJ...689.1327S}. 

Mass measurements of a planet companion of such a faint object ($V=18.5 ~mag$) would normally be outside the technical capabilities of current instrumentation; however, \tplanet has some unique properties. First, the planet/star mass ratio is large. The minimum mass scenario for TOI-263b is that of an inflated Neptune with 1 \Mnep, which would produce a radial velocity signal with an amplitude of about 35\,m/s. Larger amplitudes are expected for more massive planets. Second, the short orbital period not only makes this object unique from the formation perspective, but conspires to enlarge the RV signal and allows observations to cover more than half of the orbital period of TOI-263b during a single night. Third, despite the faintness of the primary, ESPRESSO at the VLT \citep{Pepe2014} can provide a RV precision enough to detect the minimum mass (1 \Mnep) of TOI-263b at the 5-$\sigma$ level. 


Here, we report the outcome of such observations, which determined TOI-263b to be a brown dwarf. In section 2, we detail the observing campaign with ESPRESSO. In section 3, we refine the stellar properties of TOI-263, and the radial velocity analysis is described in section 4. Finally, in section 5 we discuss the properties of TOI-263b and its possible formation history.

\section{Observations}
\label{sec:Observations}

A time series of radial velocity measurements were taken with ESPRESSO \citep{Pepe2014} on the night of 4 November 2019, under program ID 105.20ND (PI E$.$ Pall\'e). The observing plan was to monitor \tstar uninterruptedly from twilight to twilight in the 4UT mode, lasting for nearly 9 hours, and allowing us to sample 70\% of the orbital period. By carefully choosing the observing night, we arranged to cover the maximum and minimum RVs at phases 0.25 and 0.75 (the orbital period of TOI-263b is 13.5 hours). The continuous monitoring is important, as stellar activity is known to hamper the signal of planets orbiting M stars \citep{Reiners2010, Gomes2012}. By sampling continuously the orbital period we would normally be "freezing" the rotation-modulated activity of the star, as it was expected at rather long time scales. For its spectral type, a typical rotation period of $>20$ days was presumed since no obvious photometric variability had been identified in the TESS light curves \citep{Parviainen2020}. We will see, however, that a more in-depth analysis of the TESS light curve reveals the star to be a fast rotator.  

Unfortunately, our observing night was affected by adverse weather conditions and technical problems. At the beginning of the night, thick clouds prevented any data acquisition until about 1:30 UT, when the telescopes were first open. At that time, however, UT1 telescope unit failed and, after some tests, it was decided to start and continue observations during the night with 3 UTs only. Final science observations were taken from 2:10 UT through 8:48 UT and consisted of 12 consecutive spectra of 30 min of exposure time. From roughly 4:00 to 4:30 UT there is a gap in the observations as the object was crossing the meridian. The averaged S/N of each individual spectrum is $\sim$10 at 550\,nm. 

We reduced the raw spectral images using the ESPRESSO pipeline (version 2.0.0) within the \texttt{EsoReflex} environment \citep{Freudling2013}. The data were processed using standard procedures, and the extracted spectra were deblazed, slice-merged, and order-merged. During the observations, the target was on fiber A, while fiber B pointed to the sky for a simultaneous monitoring of the sky emission. The sky spectra were scaled and subtracted from the science data by the pipeline to account for the fiber-to-fiber relative efficiency. For each reduced ESPRESSO spectrum, the cross-correlation function (CCF) of the data is computed with respect to an M3 stellar mask and the radial velocity together with the full-width-half-maximum, contrast, and bisector span are obtained from a Gaussian fit to the CCF. However, due to the low S/N of the data and the high rotation of the star (see ~\ref{sec:tesslc}), the CCF computation failed to provide reliable RVs. Thus, we used the \texttt{SERVAL} code \citep{SERVAL}, which produces high-precision differential radial velocities by means of template matching, with the template constructed as the combination of our 12 spectra. To improve the RV precision, as a previous step, we masked in the observed spectrum the Hydrogen lines and several other strong chromospheric lines, as they may have different activity-induced RV values.

Using the ESPRESSO exposure time calculator in the 4UT mode, $8\times4$ binning, and assuming an airmass of 1.3, a seeing of 1.2\arcsec{}, and an M3 stellar template, a radial velocity precision of around 5\,m/s in 1 hour exposure time was expected. Our retrieved RV values, however, have a mean precision of  $\sim$160\,m/s for a 30\,min exposure time, which is about a factor of 30 worse than expected. The difference is mainly attributed to the fast rotation of the star, the use of 3 UTs instead of 4 UTs, and last, but not least, the rapid orbital motion of all components that is noticeable in 30 min exposures.  The individual RV measurements and their errors are given in Table~\ref{tab:rvtab}.

\begin{table}
    \centering
    \small
    \caption{Derived ESPRESSO RVs and \halpha pEWs for TOI-263b.} \label{tab:rvtab}
    \begin{tabular}{@{}clrc@{}}
        \hline\hline
        \noalign{\smallskip}
        Spectrum & \multicolumn{1}{c}{BJD} & \multicolumn{1}{c}{RV}  & \multicolumn{1}{c}{pEW H$\alpha$}\\ 
                 &            & \multicolumn{1}{c}{($\mathrm{km\,s^{-1}}$)} & \multicolumn{1}{c}{(\AA)} \\
        \noalign{\smallskip}
        \hline
        \noalign{\smallskip}
1  & 2458791.609465 &  $-$17.02$\pm$0.22 &  $-$5.60$\pm$0.50 \\  
2  & 2458791.631652 &  $-$19.70$\pm$0.16 &  $-$4.80$\pm$0.50 \\  
3  & 2458791.648226 &  $-$20.95$\pm$0.18 &  $-$4.51$\pm$0.50 \\  
4  & 2458791.667654 &  $-$21.80$\pm$0.17 &  $-$4.35$\pm$0.50 \\  
5  & 2458791.710933 &  $-$19.25$\pm$0.12 &  $-$4.17$\pm$0.50 \\  
6  & 2458791.733222 &  $-$15.57$\pm$0.17 &  $-$4.20$\pm$0.50 \\  
7  & 2458791.757235 &  $-$10.33$\pm$0.22 &  $-$4.08$\pm$0.50 \\  
8  & 2458791.777535 &   $-$4.11$\pm$0.18 &  $-$3.87$\pm$0.50 \\  
9  & 2458791.797632 &   $+$1.81$\pm$0.12 &  $-$3.55$\pm$0.50 \\  
10 & 2458791.819532 &   $+$8.36$\pm$0.14 &  $-$3.80$\pm$0.50 \\  
11 & 2458791.841017 &  $+$13.23$\pm$0.14 &  $-$4.12$\pm$0.50 \\  
12 & 2458791.864900 &  $+$20.36$\pm$0.15 &  $-$4.48$\pm$0.50 \\  
        \noalign{\smallskip}         
        \hline
    \end{tabular}
\end{table}

\section{Stellar Properties}
\label{sec:properties}

\subsection{Distance and luminosity}
TOI-263 is detected by the Gaia survey \citep{gaia2016} at a distance of 279.4 $\pm$ 7.9 pc with colors compatible within one subtype with the M3.5 spectral typing published in \citet{Parviainen2020}. Using the Virtual Observatory tool VOSA \citep{bayo2008}, we derived the bolometric luminosity of TOI-263 by integrating over the photometric spectral energy distribution (SED) covering from the PAN-STARRS1 $g$ magnitude at 485 nm through {\sl WISE} $W2$ band at 4.6 $\mu$m. The broad-band photometry employed in this process comes from the PAN-STARRS1 \citep{flewelling2020}, 2MASS \citep{skrutskie2006}, and {\sl WISE} \citep{wright2010} surveys. Table~\ref{tab:params} provides all photometric magnitudes extracted from the on-line catalogs together with the Gaia distance and proper motion. To complete the photometric SED towards bluer and redder wavelengths we used a BT-Settl synthetic spectrum \citep{allard12,baraffe17} with an effective temperature $T_{\rm eff}$ of 3400 K, solar metallicity and surface gravity log\,$g$ of 5.0 dex (see subsection \ref{sec:spec}). This model actually provides a very nice reproduction of the observed photometric SED at all photometric bands according to the VOSA fitting algorithm. The resulting bolometric luminosity and its error bar are listed in Table~\ref{tab:params}. Given the mass of the brown dwarf companion (Section~\ref{sec:Radvel}), its contribution to the total luminosity of the TOI-263 system is expected to be of the order of a factor of 100 smaller than that of the parent star. 
\begin{figure}
	\centering
	\includegraphics[width=\columnwidth]{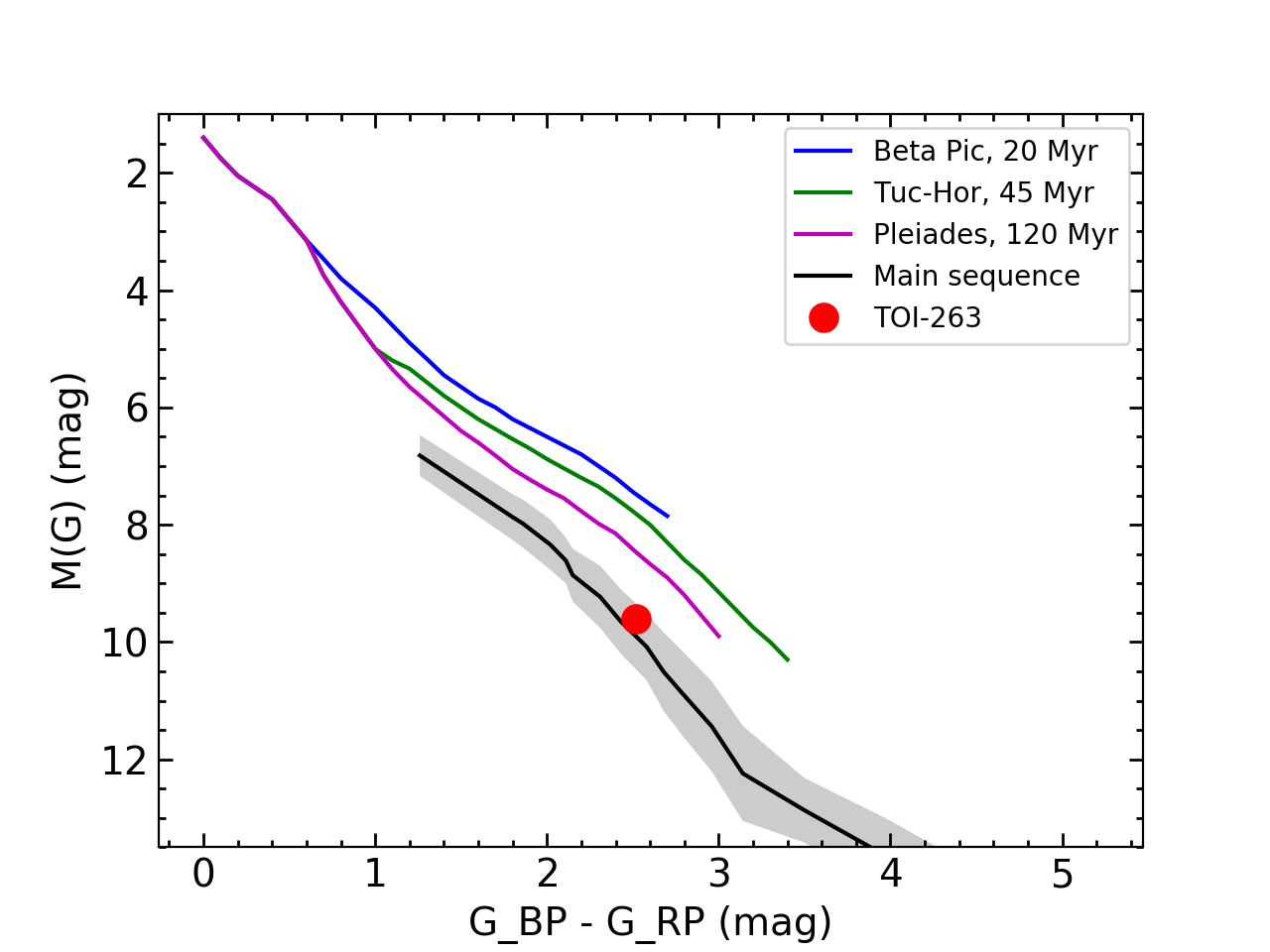}
	\caption{TOI-263 is located in the Gaia color-magnitude diagram together with the mean sequences of young clusters and moving groups \citep{luhman18} and the main sequence of stars \citep{cifuentes20}. From its location, we infer that the age of TOI-263 is close to that of the main sequence of stars near the Sun. The error bars of TOI-263 are smaller than the symbol size. The gray area represents the 1-$\sigma$ dispersion of field M dwarfs.
	}
	\label{fig:colormag}
\end{figure}

Using the accurate photometry and distance provided by Gaia, we constructed the color-magnitude diagram shown in Fig.~\ref{fig:colormag}, where we also depict the well known, empirically determined mean sequences of stellar members of the $\beta$ Pictoris moving group ($\sim$20 Myr, \citealt{miret-roig20}), Tucana-Horologium moving group ($\sim$45 Myr, \citealt{bell15}), the Pleiades open cluster ($\sim$120 Myr, \citealt{gossage18}), and the field (possible ages in the range 0.8--10 Gyr). These sequences were taken from \citet{luhman18} and \citet{cifuentes20} and were derived by employing Gaia data; therefore, the direct comparison with TOI-263 is feasible without any systematic effect. From its location in the Gaia color-magnitude diagram, we infer that TOI-263 has a likely age near that of the "field" and is not as young as the Pleiades cluster. We did not correct TOI-263 data for interstellar extinction because from its optical and infrared photometry and optical spectroscopy there is no evidence of strong or anomalous absorption. The "field" age is consistent with the stellar lithium depletion and the measured mass and radius of the brown dwarf companion (see section~\ref{sec:Radvel}). 

The mass and radius of the parent, early-M star were updated using the mass---$K$-band magnitude and the mass---radius relationships defined in \citet{mann19} and \citet{Schweitzer2019}. These relations, which are valid for field M-type stars with near-solar and solar metallicity, are widely used in the exoplanetary field. The absolute $K$-band magnitude of TOI-263 is 6.015 $\pm$ 0.100 mag, where the error bar accounts for the uncertainties in the photometry and trigonometric parallax. The derive values are compatible with those of \citet{Parviainen2020} within 1-$\sigma$ the quoted uncertainties and are listed in Table~\ref{tab:params}. The updated stellar mass and radius determinations are employed throughout this work. 

\subsection{TESS photometric analysis}
\label{sec:tesslc}

\begin{figure}
	\centering
	\includegraphics[width=\columnwidth]{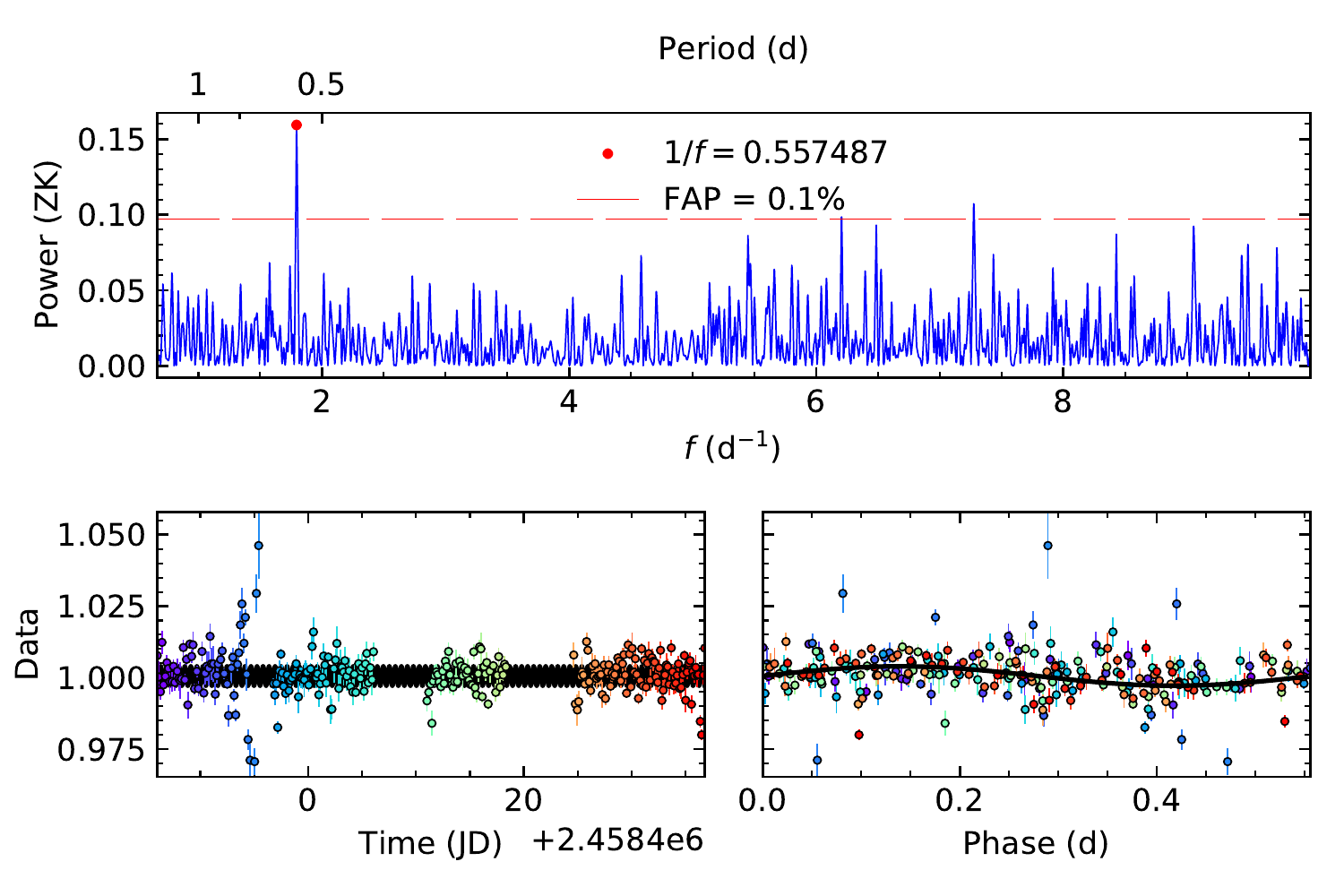}
	\caption{Top: GLS periodogram of the TESS light curve from Sectors 3 and 4, (binned in 3\,h intervals). The red dashed line indicates a false alarm probability threshold of 10\%. The highest peak in the periodogram marked with a red dot corresponds to a period of 0.5575$\pm$0.0008\,d. Bottom left: TESS photometric time series modelled with a sinusoidal with a period corresponding to the highest peak in the GLS periodogram above. Bottom right: phase-folded TESS photometry to the aforementioned period, in days.
	}
	\label{fig:tesslc}
\end{figure}

\begin{figure}
	\centering
	\includegraphics[width=\columnwidth]{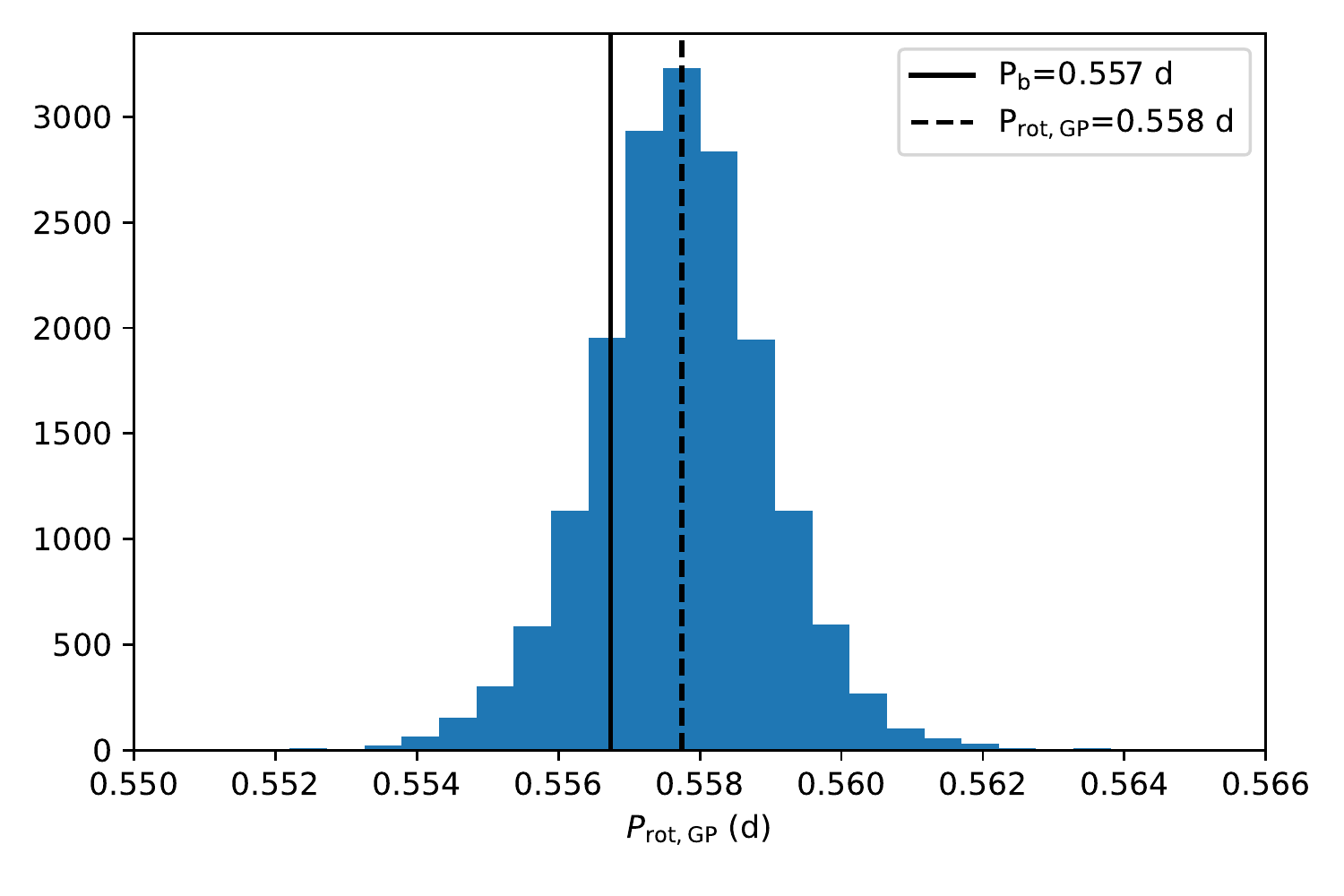}
	\caption{Posterior distribution function of the $P_\textnormal{rot}$ hyperparameter of the GP kernel. The vertical dashed line indicates the median of the distribution, while the solid vertical line indicates the orbital period of the substellar companion TOI-263~b.
	}
	\label{fig:tesslc2}
\end{figure}


To explore the rotational period of the star, we use the Sector 3 and 4 \textit{TESS} photometric light curves extracted from the full frame images using \texttt{eleanor} \citep{eleanor}. We did not use the \textit{TESS} photometry provided by the Science Processing Operations Center \citep[SPOC;][]{SPOC} at the NASA Ames Research Center, since this target was only observed in 2-min cadence during Sector 3, and not in Sector 4. 

We first checked for variability in the light curves from both Sectors with a generalized Lomb-Scargle periodogram \citep[GLS;][]{GLS}. There is a highly significant peak (FAP$<$0.1\%) with a period of $0.557\pm0.001$\,d in the original 30-min cadence data. The same, highly significant period is found when binning the data into 5\,h intervals to reduce the short-time variability (Fig.~\ref{fig:tesslc}, top panel). Modelling the photometric variability detected with the GLS periodogram with a sinusoid we measure the amplitude of such signal to be $3.4\pm0.4\,\mathrm{ppt}$ and $2.9\pm0.6\,\mathrm{ppt}$ for the 30-min original cadence photometry and the 3-h binned data, respectively (Figure~\ref{fig:tesslc}). Moreover, we model the \textit{TESS }photometry as in \citet{Luque2019} with a quasi-periodic Gaussian Process (GP) kernel as defined in \citet{celerite} of the form 
\begin{equation*}
k_{i,j}(\tau) = \frac{B}{2+C}e^{-\tau/L}\left[\cos \left(\frac{2\pi \tau}{P_\textnormal{rot}}\right) + (1+C)\right] \quad ,
\end{equation*}
where $\tau = |t_{i} - t_{j}|$ is the time-lag, $B$ and $C$ define the amplitude of the GP, $L$ is a timescale for the amplitude-modulation of the GP, and $P_\textnormal{rot}$ is the rotational period of the QP modulations. The posterior samples of $P_\textnormal{rot}$ are shown in Fig.~\ref{fig:tesslc2}, measuring a $P_\textnormal{rot}=0.557\pm0.001$\,d as detected from the GLS analysis. 

Considering that the orbital period of the companion TOI-263~b is $0.5568140\pm4.1\times10^{-6}$\,d as computed by \citet{Parviainen2020}, we checked that the measured variability does not correspond to the transits of the planet. Thus, we masked the transits and carried out the previous analyses again, finding the same results. In addition, we check the Sector 3 \textit{TESS} photometry provided by SPOC and we find the same results, for both simple aperture photometry or systematics-corrected photometry \citep[PDC,][]{Smith2012PASP..124.1000S, Stumpe2012PASP..124..985S, Stumpe2014PASP..126..100S}.

We also attempted to model the phase variability using a physical phase curve model including reflection, thermal emission, Doppler boosting, and ellipsoidal variations with \texttt{PyTransit} \citep{Parviainen2015}. However, the amplitude of the variations is significantly larger than what would be expected and the shape of the phase curve is not modelled well with these standard components. Each of the phase curve components is expected to have an amplitude in the range of 500--1000~ppm, what is significantly lower than what can be reliably measured over a companion orbit. 


\subsection{Spectral analysis}
\label{sec:spec}


\begin{figure}
	\centering
	\includegraphics[width=\columnwidth]{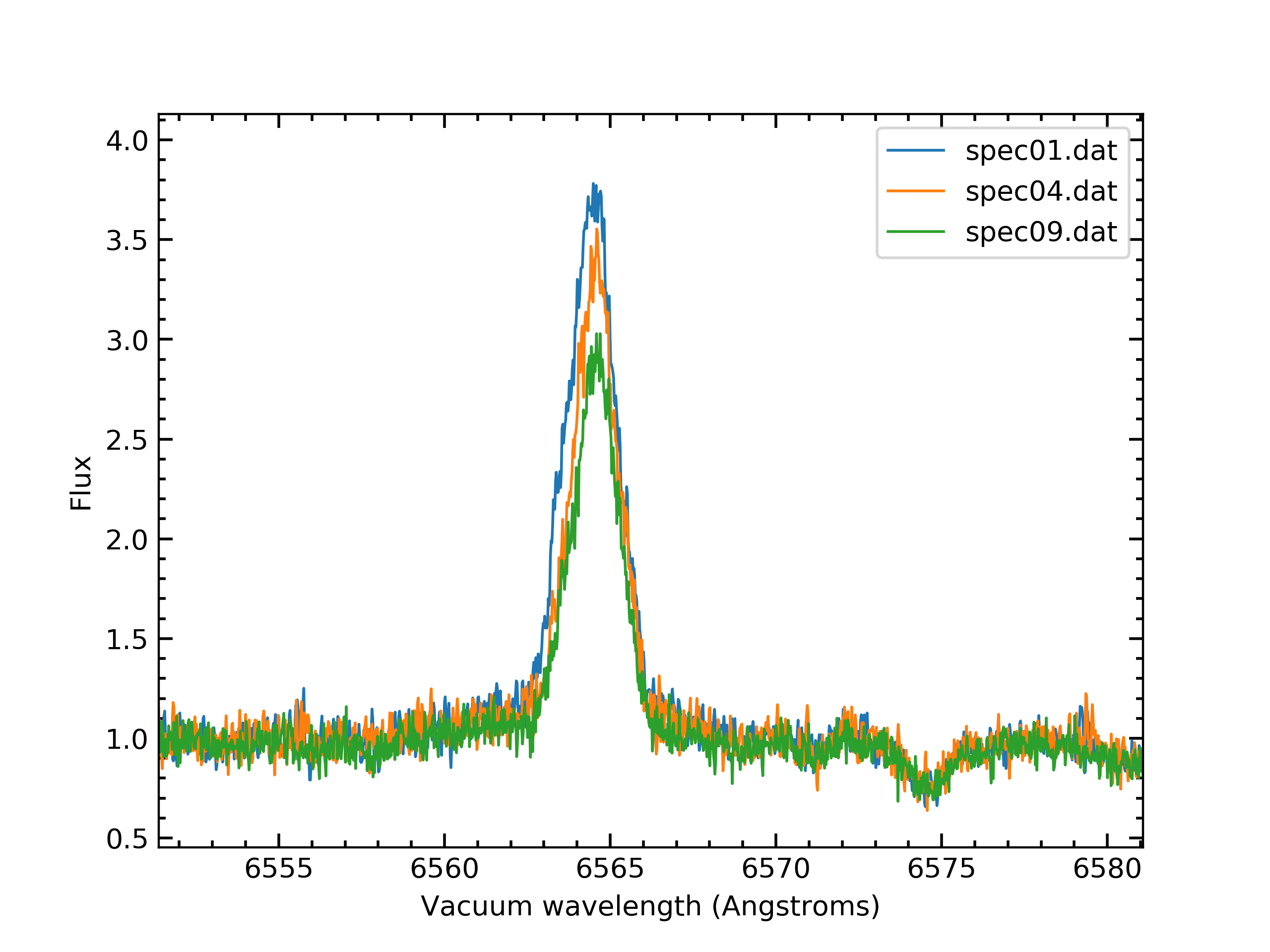}
	\includegraphics[width=\columnwidth]{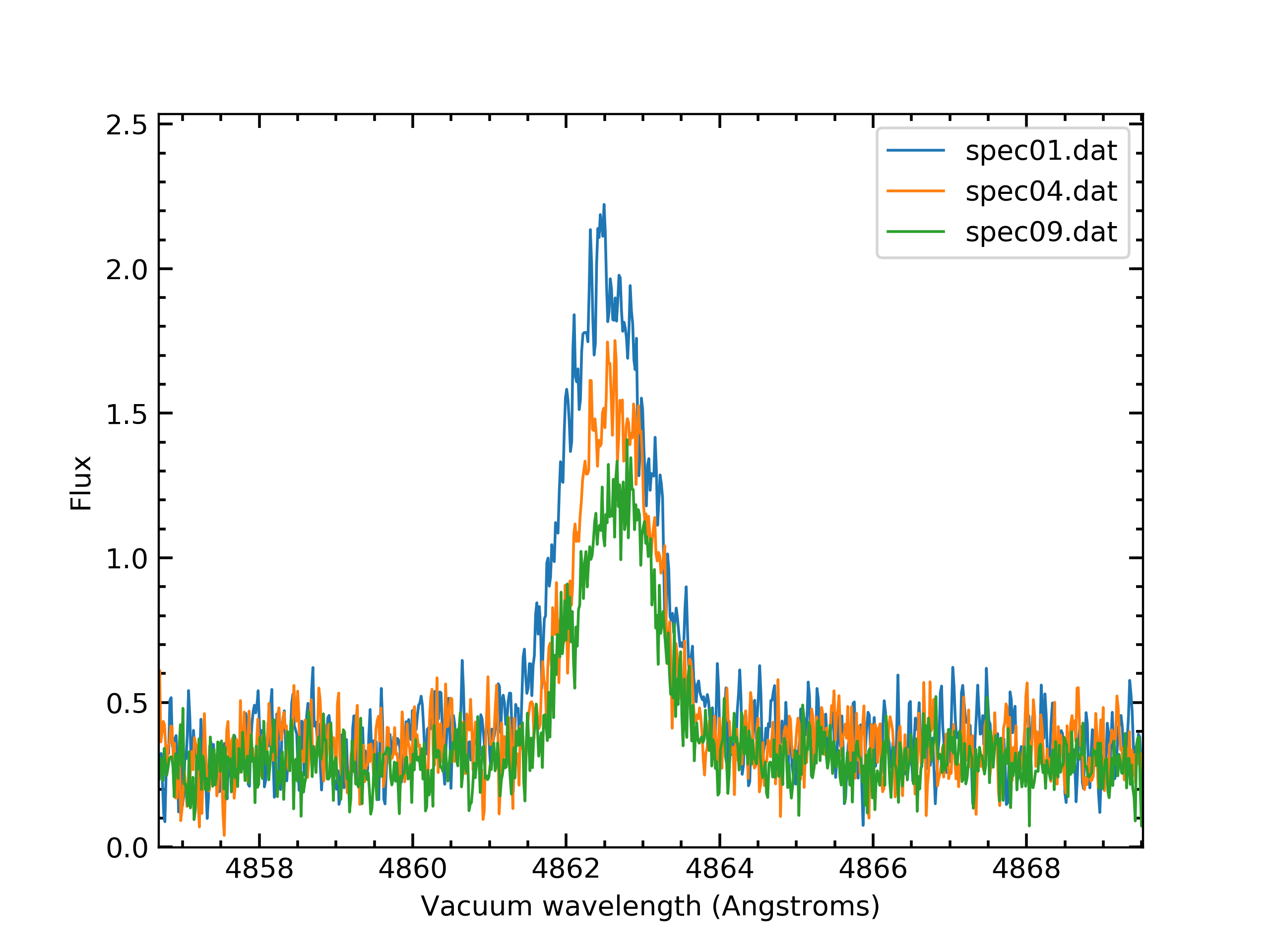}
	\caption{A zoom of the TOI-263 ESPRESSO spectra normalized to the pseudo-continuum in the spectral regions of $H_\alpha$ (top panel) and $H_\beta$ (lower panel). Variability in both spectral lines at hourly time scales is visible in the spectrum. 
	}
	\label{fig:hlines}
\end{figure}

In order to determine the spectroscopic rotational velocity of TOI-263, we corrected all ESPRESSO spectra for the different velocities and combined all Doppler-shifted data into one single spectrum, thus increasing the S/N considerably. Then, we used a “non-rotator” of the same spectral type (using the same spectral type is important to break the degeneracy between temperature, gravity and $v$\,sin\,$i$ in the cool dwarf domain), observed with ESPRESSO and CARMENES, to derive $v$\,sin\,$i$ = 38.6\,$\pm$\,0.6 km\,s$^{-1}$. This $v$\,sin\,$i$ value agrees with the fast rotation period determined from the TESS photometry further confirming that the rotational period of the star and the orbital period of the planet are synchronized. According to the ESPRESSO data, TOI-263 is indeed a fast rotator in marked contrast with what is expected for its spectral type and age \citep{Zapatero2006, Curtis2020}. 

A detailed analysis of the individual ESPRESSO spectra reveals several interesting facts. First, there is no strong lithium in the stellar spectrum (upper limit on pseudo-equivalent width pEW =  40 m\AA), indicating that the star has severely depleted lithium, and thus an age > 20 Myr, which agrees with the dating inferred from the Gaia color-magnitude diagram of Fig.~\ref{fig:colormag}.  
Second, the Balmer lines are persistent in all spectra, are seen in emission, and they are variable (see Figure~\ref{fig:hlines} for the $H_{\alpha}$ and $H_{\beta}$). Some line profiles are not symmetric, indicative of the presence of two or more components such as rotation, flare activity, chromospheric variability and/or contributions from the companion. TOI-263 is not detected by GALEX or Rosat, preventing us to explore the existence or not of UV excesses. We measured the pseudo-equivalent width (pEW) of \halpha emission by integrating the line profile between 655.9 and 656.9 nm; error bars were determined by changing the pseudo-continuum within the dispersion of the observed flux. The obtained values are given in Table~\ref{tab:rvtab} and shown as a function of time in Fig.~\ref{fig:rvs}. Thus, \tstar is a chromospherically active star with variations of \halpha emission \citep{Reid1995}.



Finally, we used the automatic tool {\sc SteParSyn} \citep{Tabernero2018,Tabernero2020b} to infer the stellar atmospheric parameters based on the new ESPRESSO data ($T_{\rm eff}$, log(g), see Table~\ref{tab:params}). The latest version of {\sc SteParSyn} relies on {\tt emcee} \citep{emcee}, a Markov Chain Monte Carlo (MCMC) method used to fully sample the underlying distribution of the stellar atmospheric parameters. We computed a synthetic grid using the PHOENIX BT-Settl \citep{Allard2011} model that are particularly tailored to model low-mass late-K and M dwarfs alongside the radiative transfer code {\tt turbospectrum} \citep{Plez2012}. The line list was collected using the VALD\footnote{http://vald.astro.uu.se/} and \citet{Plez2012} databases. Our modelling rests on the fit to the TiO band system at $7\,050\:$\AA. Stellar metallicity was fixed to solar. From spectral fitting we obtain a $T_{\rm eff}$ = $3471 \pm 33$~K, $\log{g}$ = $4.67 \pm 0.03$~dex, and  $v$\,sin\,$i$ = $37.92 \pm 0.36$ km\,s$^{-1}$, the former being compatible within 1 $\sigma$ with that obtained previously from line broadening. These values can be found in Table~\ref{tab:params}.

\section{Radial Velocity Analysis}
\label{sec:Radvel}

\begin{figure}
	\centering
	\includegraphics[width=\columnwidth]{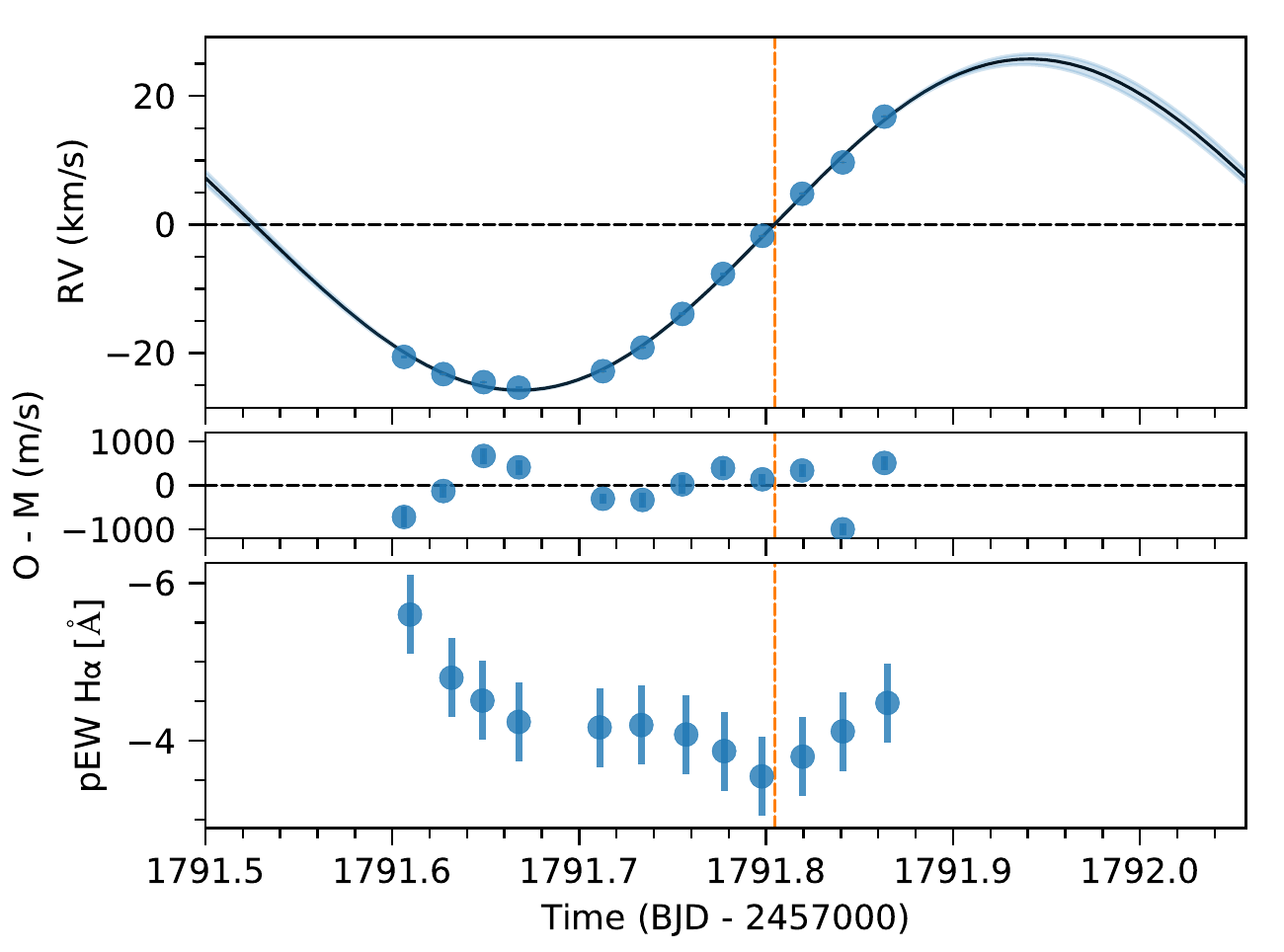}
	\caption{Top panel: Individual ESPRESSO-3UT RVs of TOI-263 are shown with blue circles. The solid line represents the best Keplerian fit to the RVs, after setting strong priors on the orbital period and the phase of the brown dwarf from TESS and ground-based transit photometry analyses. Error bars fall within the size of the symbols. Middle panel: RV residuals between the data and the best fitting model. The std of the RV residuals is 490 $m/s$ and show no significant periodicity. Bottom panel: pseudo-equivalent width of H$\alpha$ emission of TOI-263. The minimum occurs when the brown dwarf is hidden by its parent star ("secondary transit"), marked with a broken orange line.
	}
	\label{fig:rvs}
\end{figure}

Figure~\ref{fig:rvs} shows the ESPRESSO RVs measured for \tstar, which in our case are both a time series and phase-folded measurements. In the figure, the data are plotted versus the observing date.

We used {\sc Exo-Striker} \citep{exostriker} to determine the orbital parameters of \tplanet. The negative logarithm of the likelihood function ($-\ln\mathcal{L}$) of the model is minimized while optimizing the planet orbital parameters, i.e., RV amplitude $K$, eccentricity $e$, argument of periastron $\omega$, and a RV zero-point offset; but setting strong normal priors in the period and the mid-transit time from the results of \citet{Parviainen2020}. Afterward, we estimate the uncertainties of the best-fit parameters using the Markov Chain Monte Carlo (MCMC) sampler \texttt{emcee} \citep{emcee}. We adopted flat priors for all fitted parameters and selected the 68.3 confidence interval levels of the posterior distributions as $1\sigma$ uncertainties. The {\tt serval} pipeline also computes several indices that can be used as stellar activity proxies \citep{SERVAL}. There is a positive correlation (p=0.62) between the H$\alpha$ pEW and the dLW indices measured with SERVAL, but there are no correlations with the remaining activity indicators such as the CRX or the NaD indices. The RV residuals, on the other hand, correlate with the CRX index (p=-0.63). Such a negative correlation is expected for active M dwarfs, and it is indicative that the remaining variability in the RVs can be attributed to stellar activity induced by spots coupled with the rotation of the star \citep[e.g.,][]{Lev2018,Baroch2020A&A...641A..69B}.

We do observe, however, that the pseudo-equivalent width of H$\alpha$ emission has a minimum during the secondary eclipse of \tplanet (see Figure~\ref{fig:rvs}). We speculate that this could reflect the interaction between the magnetic fields of the star and the brown dwarf, where part of the H$\alpha$ emission comes from the sub-brown dwarf stellar point or an arc of mass transfer between the star and the brown dwarf, which is occulted during the secondary eclipse. Emission from the surface of the companion itself cannot be ruled out either.


Our best fit results yield a Keplerian solution with an amplitude of $K=25.74^{+0.47}_{-0.42}\,\mathrm{km\,s^{-1}}$ and very small eccentricity ($e=0.02 \pm 0.01$) suggesting that the brown dwarf orbit is mostly circular. Assuming the stellar parameters from \citet{Parviainen2020} ($M_\star=0.4 \pm 0.1\,M_\odot$, $R_\star=0.405 \pm 0.077\,R_\odot$), this implies that \tplanet is a brown dwarf with a mass of  $61.6 \pm 4.0\,\Mjup$. The detailed derived properties for \tplanet are given in Table~\ref{tab:params}. We also derive an equilibrium temperature of $T_{eq} = 1014 K$, however, this number is only useful for comparison with other ultra-short period planetary objects. Depending on the age of the system, the intrinsic temperature of the brown dwarf, which is age dependent, could be higher. For a $\approx 60\,\Mjup$ brown dwarf the intrinsic temperatures at 1, 2, and 5 Gyr are 1800 K, 1470 K, and 1120 K, respectively \citep{baraffe17}.


\begin{figure}
	\centering
	\includegraphics[width=1.1\columnwidth]{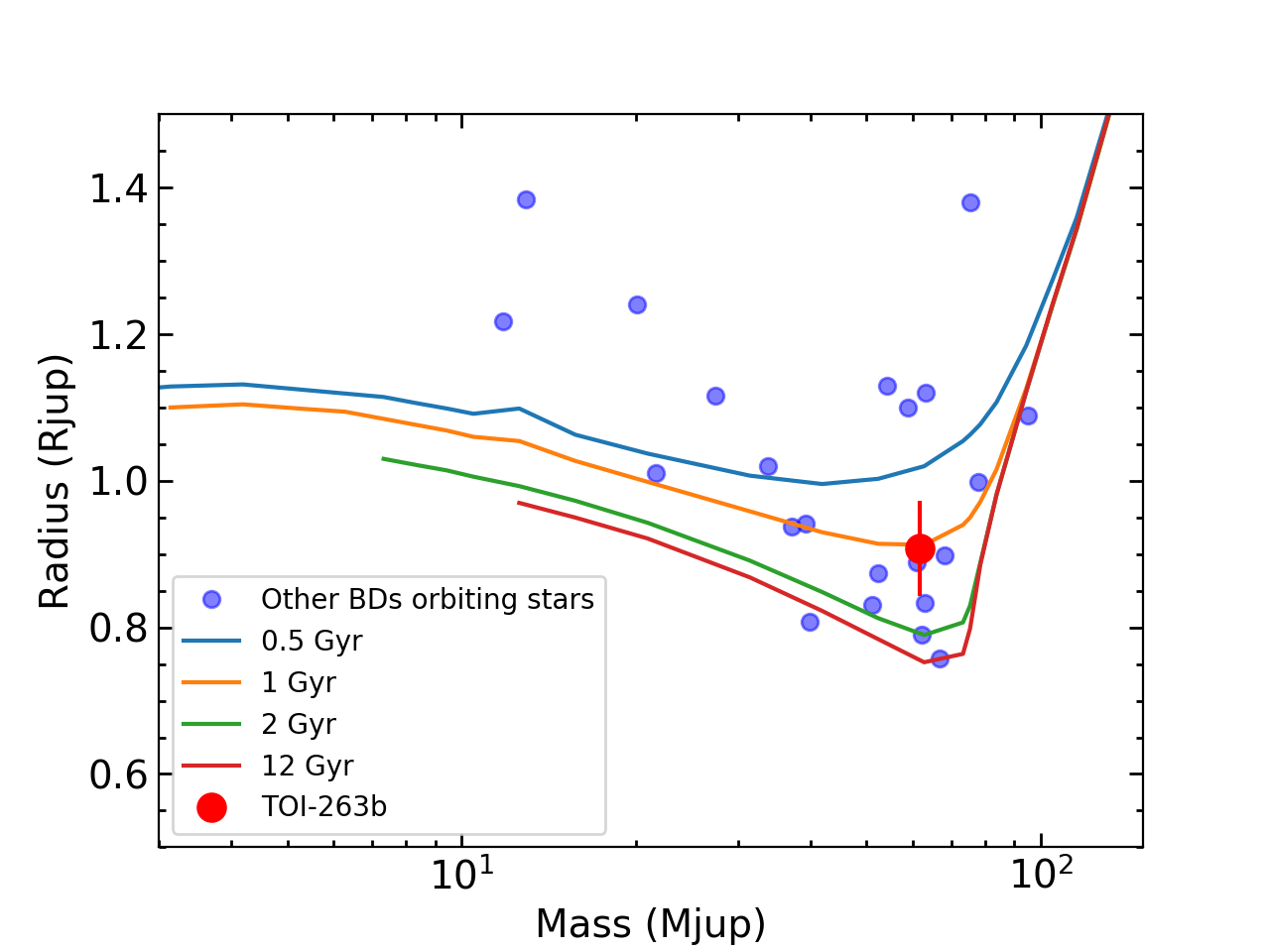}
	\caption{Mass--radius diagram for all brown dwarfs found orbiting stars. TOI-263\,b is shown with the solid, red circle. The BT-Settl isochrones by \citet{allard12} and \citet{baraffe17} are also shown. }
	\label{fig:evol}
\end{figure}

\begin{table*}
	\centering
	\normalsize
	\caption{Stellar and Brown dwarf parameters.}
	\begin{tabular}{@{}lllc@{}}
		\toprule\toprule
		Quantity & Notation & Unit & Value \\
		\midrule
		\noalign{\smallskip}
		\multicolumn{4}{l}{\emph{~~~~~Photometric and astrometric properties of TOI-263}} \\
		\noalign{\smallskip}
		Gaia photometry (Vega) & $G$ & mag & 16.840 $\pm$ 0.001 \\
		Gaia photometry (Vega) & $B_p-R_p$ & mag & 2.516 $\pm$ 0.027 \\
		Pan-STARRS1 photometry (AB) & PS1.$g$ & mag & 18.53 $\pm$ 0.50 \\
		Pan-STARRS1 photometry (AB) & PS1.$r$ & mag & 17.346 $\pm$ 0.008 \\
		Pan-STARRS1 photometry (AB) & PS1.$i$ & mag & 16.120 $\pm$ 0.004 \\
		Pan-STARRS1 photometry (AB) & PS1.$z$ & mag & 15.561 $\pm$ 0.005 \\
		Pan-STARRS1 photometry (AB) & PS1.$y$ & mag & 15.274 $\pm$ 0.005 \\
		2MASS photometry (Vega) & $J$ & mag & 14.078 $\pm$ 0.030 \\
		2MASS photometry  (Vega) & $H$ & mag & 13.450 $\pm$ 0.038 \\
		2MASS photometry (Vega)  & $K$ & mag & 13.246 $\pm$ 0.040 \\
		WISE photometry  (Vega) & $W1$ & mag & 13.151 $\pm$ 0.025 \\
		WISE photometry (Vega)  & $W2$ & mag & 12.989 $\pm$ 0.026 \\
        Gaia distance & $d$ & pc & 279.4 $\pm$ 7.9 \\
		Gaia proper motion & $\mu_\alpha$\,cos\,$\delta$ & mas\,yr$^{-1}$ & 30.33 $\pm$ 0.18\\
		Gaia proper motion & $\mu_\delta$ & mas\,yr$^{-1}$ & 10.14 $\pm$ 0.18\\
		Bolometric luminosity & $L$ & L$_{\rm sol}$ & (2.716 $\pm$ 0.256) $\times 10^{-2}$ \\ 
		Effective temperature & $T_{\rm eff}$ & K & $3471 \pm 33$ \\
		Surface gravity & log\,$g$ & dex & $4.67 \pm 0.03$ \\
		Spectroscopic rotation  & $v$\,sin\,$i$ & km\,s$^{-1}$ & $37.92 \pm 0.36$ \\
		Mass of the star & $M$ & M$_\odot$ & 0.438 $\pm$ 0.036 \\
		Radius of the star & $R$ & R$_\odot$ & 0.438 $\pm$ 0.028 \\
		Age & $\tau$ &  & "field" \\
		\noalign{\smallskip}
		\multicolumn{4}{l}{\emph{~~~~~Orbital Parameters of the brown dwarf TOI-263\,b}} \\
		\noalign{\smallskip}
		Transit epoch & $T_0$ & BJD & \tzeropone \\
		Orbital period & $P$ & d & \Ppone \\
		Eccentricity & $e$ & & \ecc \\
		Semi-major axis &$a$ & au & \semimajorau \\
		Gravitational acceleration & g & $\rm m\,s^{-2}$ & \gplanet \\
		RV amplitude & $K$ & $\rm m\,s^{-1}$ & \Kp \\
		Argument of periastron & $\omega$ & deg & \omegab \\ 
		Inclination & $i$ & deg & $87.0 \pm 1.6$ \\
		\noalign{\smallskip}
		\multicolumn{4}{l}{\emph{~~~~~Properties of the brown dwarf TOI-263\,b}} \\
		\noalign{\smallskip}
		Planet radius & $R_p$ & R$_{\rm Jup}$  &  \radiusrj \\
		Planet mass & $M_p$  & M$_{\rm Jup}$  & \massMj \\
		Mean density & $\rho$ & g\,cm$^{-3}$ & \rhogcm\\
		\bottomrule       
	\end{tabular}
	\label{tab:params}  
\end{table*}

\begin{figure*}
	\centering
	\includegraphics[width=\textwidth]{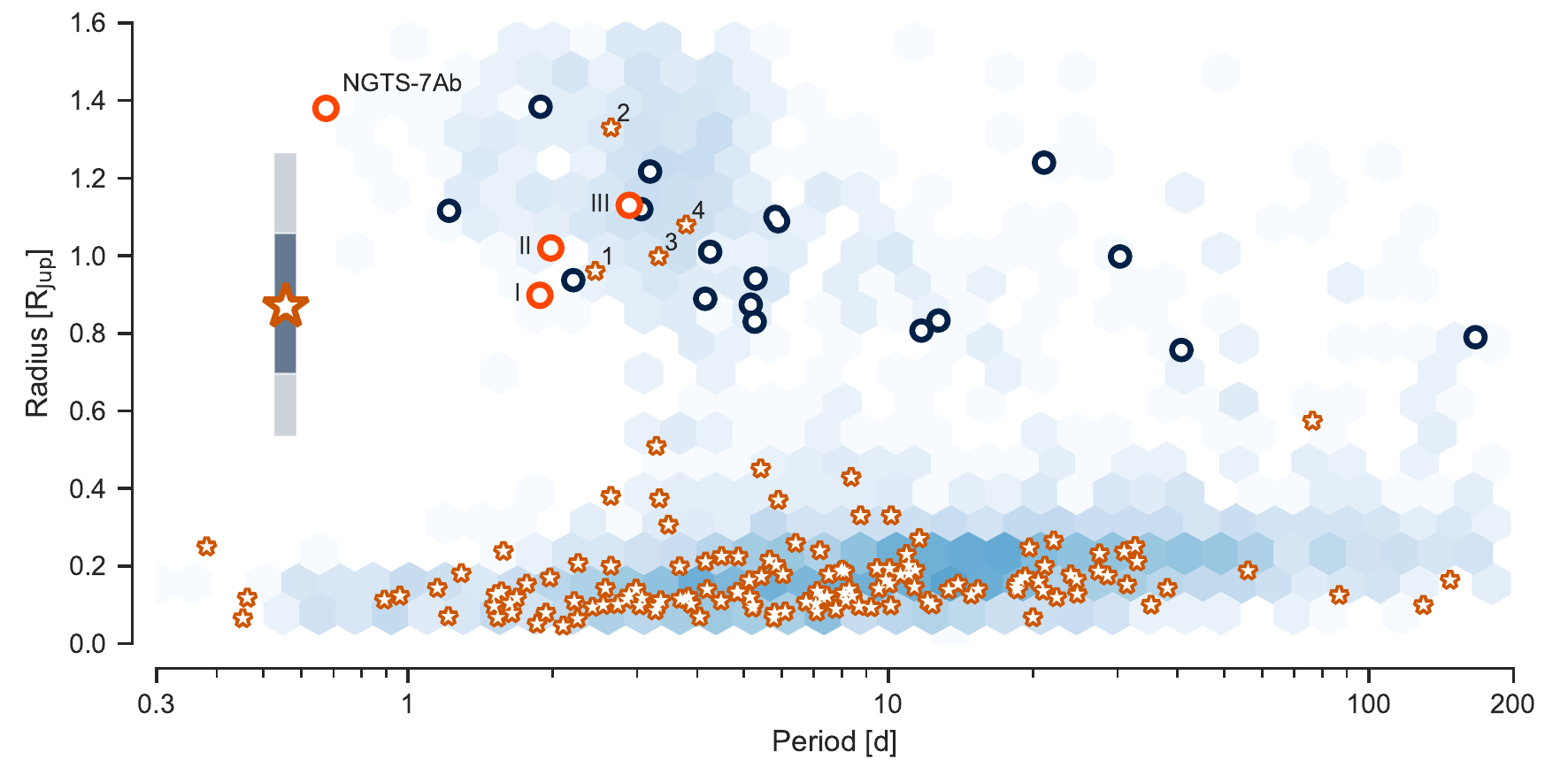}
	\includegraphics[width=\textwidth]{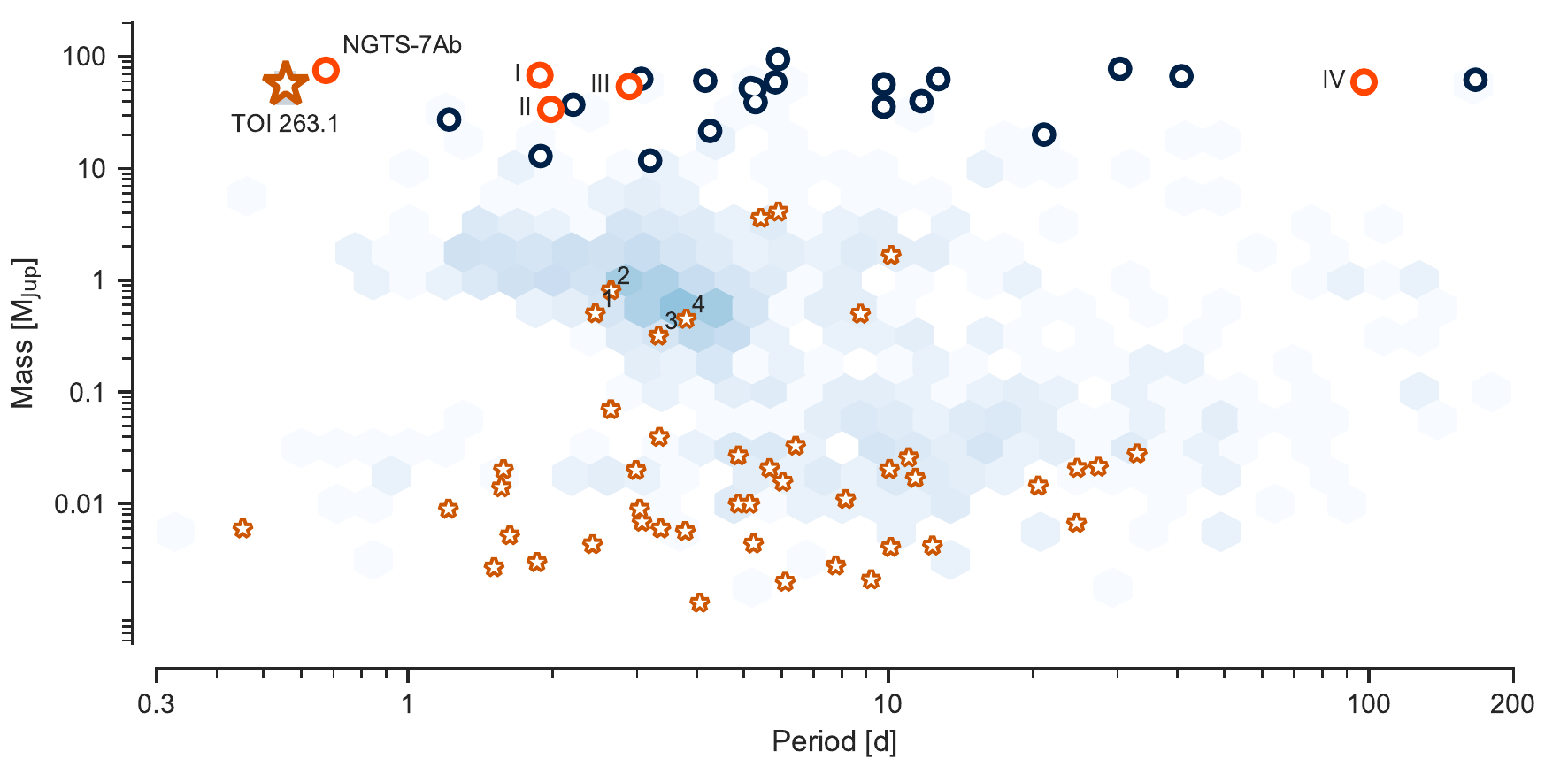}
	\caption{A period-radius ({\it top}) and period-mass ({\it bottom}) diagram of all confirmed extrasolar planets and brown dwarfs to date via transits or RVs\protect\footnotemark. The blue shaded scale indicates the density of known planets. All planets and brown dwarfs around host stars with $\teff < 4000$\,K are individually marked with a red star or a red circle, respectively. Transiting brown dwarf around other stellar types are marked with black circles. The position of \tplanet is marked with a larger red star, together with he radius and mass measurement uncertainties (within the symbol size in the latter case). For comparison, the transiting brown dwarf NGTS-7Ab is marked separately, some individual planets are marked with Latin numbers: Kepler-45 (0), NGTS-1 (1), HATS-6 (2), HATS-71A (3), and known brown dwarfs in roman numbers: LP 261-75b (I), AD 3116 b (II), NLTT 41135 b (III), RIK 72 b (IV). 
}
	\label{fig:context}
\end{figure*}

Given the small orbit of the brown dwarf around its parent M-type star, the brown dwarf likely resides in a tidally-locked orbit; furthermore, the star also rotates with the same periodicity as the brown dwarf. Therefore, any atmospheric activity-induced variation in the observed radial velocities will probably be in phase with the Keplerian signal. Indeed, there are signs of this scenario in the observed variability of the H$\alpha$ emission line, but we estimate that the effects will be smaller than the measured radial velocity amplitude ($<0.6$ km\,s$^{-1}$, \citet{Lev2018}), more so as we already masked many of the chromospherically active lines when computing the relative velocities. Note, however, that any correction for the activity-induced signal in the velocities would only diminish the mass of \tplanet, making it a more extreme inhabitant of the brown dwarf desert.

\section{Discussion}
\label{sec:disc}
\footnotetext{\url{http://exoplanet.eu/}}

\subsection{\tplanet in context}
\label{sec:context}

TOI-263\,b, with an estimated mass of $61.6 \pm 4.0\,\Mjup$ and a ultra-short period of 0.56~days, is a unique and extreme brown dwarf orbiting in the so-called ``brown dwarf desert", where there is a paucity of brown dwarfs at short orbital periods ($\lesssim$ 100~days), which contrasts with the more abundant population of close-in giant planets, the so-called "hot Jupiters" \citep{Marcy2000}.  Since the identification of the brown dwarf desert, several tens of brown dwarfs have been found by radial velocity and transit methods at close orbital separations of their stars \citep[][and reference therein]{Deleuil2008, Moutou2013, Csizmadia2015, Persson2019}. In particular, CoRoT-3\,b was the first transiting brown dwarf with a radius determination \citep[1.01$\pm$0.07 R$_{\rm Jup}$][]{Deleuil2008}, confirming the radius dichotomy between very-low mass stars, brown dwarfs and Jovian planets. All of them, together with TOI-263\,b, are depicted in the mass--radius diagram of Fig.~\ref{fig:evol}, where TOI-263\,b does not occupy any special position, and it is consistent with the BT-Settl evolutionary models for ages above 500 Myr.

Most of these brown dwarfs have been found around solar-type or more massive stars and all of them have orbital periods longer than one day, between 1.7\,d and 167\,d. 
Around cooler spectral types, however, 
there have been only five brown dwarfs with measured radii and masses.
\citep{Stassun2007, Irwin2010, Johnson2010, Gillen2017, Irwin2018, Jackman2019}. 
Most of these systems are young, 
belong to star forming associations such as Orion, USco, the Argus Young Moving Group or the Praesepe open cluster, and 
are formed by a very low-mass star and a brown dwarf or a brown dwarf binary. 
Since substellar objects cool and get fainter with time, it is very difficult to identify these double-lined eclipsing binaries at older ages like that of our Sun. 
\tplanet is the sixth  
transiting brown dwarf discovered around an M dwarf, and the one with the shortest orbital period, a record previously hold by NGTS-7Ab \citep{Jackman2019}. In fact, TOI-263\,b and NGTS-7Ab are the only known brown dwarfs with orbital periods shorter than one day.

Figure~\ref{fig:context} shows period--radius and period--mass diagrams of all confirmed extrasolar planets and brown dwarfs to date discovered via transit, and with radial velocity measurements, putting \tplanet in context. 
The period--radius diagram shows \tplanet to be completely isolated, at a bit more than half the radius of NGTS-7Ab. Other brown dwarfs with similar radii have much larger orbital periods (a factor 4 or larger). 
In the period--mass diagram, \tplanet is much more similar to NGTS-7Ab, and has an average mass regarding the population of known transiting brown dwarfs. 

\subsection{Formation Scenarios, Synchronization and Tidal decay}
\label{sec:formation}

The existence of a brown dwarf at an orbital period of 0.56 days around an M3.5 dwarf is challenging in terms of determining its formation history. 
Also, given that spin--orbit synchronization has been already achieved, we have no way of knowing its dynamical history. Nevertheless, the high companion-to-stellar mass ratio, $q$, of 0.09--0.21 suggests that fragmentation scenarios are more likely than core accretion scenarios because the former processes, if possible, occur much more efficiently than the latter \citep[e.g.,][]{Chabrier2014}. 

Fragmentation scenarios include simultaneous formation of close binary systems or formation of gaseous clumps via disk instability followed by orbital migration. Magnetohydrodynamic (MHD) simulations show that collapse of molecular clouds result in three different outcomes depending on the ratio of its initial rotational to magnetic energies \citep[e.g.,][]{Machida2008}: Provided an initial cloud has a large magnetic energy relative to a rotational energy, the ohmic dissipation works effectively in forming a close binary system. However simulations also show that the resultant mass ratio is not so small \citep[$q \gtrsim$ 0.3;][]{Bate2002}, suggesting that the TOI-263 system ($q < 0.21$) is unlikely to have formed like close binary systems.

On the other hand, recent hydrodynamic simulations for disk instability demonstrate that fragmentation (or clump formation) occurs around low-mass stars, when the circumstellar disk is 0.3--0.6 times as massive as the central star \citep{Mercer2020}. Such a disk mass is somewhat larger than but similar to the measured mass of \tplanet. Given that not all the initial disk gas accretes to a brown dwarf, the TOI-263 system may be explained by the disk fragmentation followed by inward migration to the current close-in orbit \citep[e.g.,][]{Forgan2018}. A key question to be solved would be whether such a close companion survives orbital decay and engulfment by its host star.

\citet{Donati2008} showed that the rotation period of the planet-hosting star $\tau\,Boo$ is synchronized with the orbital motion of its giant planet, $\tau\,Boo\,b$. Using their equation, one gets that the TOI-263 system is synchronized quite soon in its evolution, after 6 Ma. This clearly contrasts with the tau Boo system because here the companion is a brown dwarf, more massive than the Tau boo planet, and it is located closer to its parent star, which is the main driver for a faster synchronization.

A companion orbiting in synchronisation with stellar spin evolves not through tidal energy dissipation but through loss of the total angular momentum by magnetic braking \citep{Barker2009,Damiani2016}. In the case of \tplanet, we have also seen that the star is magnetically active and that the brown dwarf is in a tidally-locked orbit, suggesting that both mechanisms are likely in place. Following \citet{Damiani2016} and using the observed properties of the TOI-263 system, the orbital decay is estimated to occur on a timescale of Myr. Considering that the estimated stellar age is $>$~10-20~Myr from the depletion of lithium in the parent star and $>$~500~Myr from the brown dwarf mass-radius relationship (Fig.~\ref{fig:evol}), one would expect \tplanet to already have been engulfed by the host star, however, either a very high $Q\gtrsim10^9$ (weak tides), or very weak braking $\gamma\lesssim10^{-2}$, could explain the timescale of TOI-263b.
In any case, as \cite{Jackman2019} suggest, more observations of transiting brown dwarfs, in particular unstable systems,  are required to test evolutionary scenarios. The TOI-263 system is an excellent target for this endeavor.

Finally, \cite{Fontanive2019} studied the role of stellar multiplicity on close in massive planets and substellar companions. They conclude that binarity plays a crucial role in the existence of such objects, which are almost exclusively observed in multiple systems. Using Gaia Second Data Release \citep[DR2;][]{Gaia2018} and the 1-$\sigma$ error bars, we do not detect any other source with the same distance and proper motion as \tstar within a projected separation of radius of $10^5$ au. Thus, any hypothetical companion to \tstar, if it exists, would be either fainter than TOI-263,  or closer than 0.5''(corresponding to a projected separation of $\approx 140~au$ at TOI-263’s distance of 279 pc, and not detected by Gaia\footnote{\emph{Gaia} is complete to $G\approx17$ and limited to $G\approx21$; using PHOENIX BT-Settl models for an age of $0.5$\,Gyr \citep{Baraffe2015}, we estimate corresponding masses of $0.37$ and $0.096\mathrm{\,M}_\odot$. Hence, mid--late M dwarf companions are possible. Gaia DR2 and EDR3 data aree incomplete below 0.5''}), or located at an orbital separation larger than $10^5$ au.

\section{Conclusions}
\label{sec:Concl}

We report here ESPRESSO 3-UT mode observations of the \tstar system, and found that \tplanet is low-mass brown dwarf with a mass of $61.6 \pm 4.0 \,\Mjup$. With an ultra-short period of 0.56~days, \tplanet is the shortest period known brown dwarf around any stellar type, and a unique and extreme inhabitant of the so-called ``brown dwarf desert". It is also one of the lightest among transiting brown dwarfs found so far around stars of cold ($\teff < 4000$\,K) stellar type. We found that the orbital period of \tplanet is synchronized to the rotation period of the host star, indicative of interaction between the star and the brown dwarf than spun the stellar rotation, and that the star is relatively active. All these findings combined, suggest that the system formation history might be explained via disc fragmentation and later migration to close-in orbits or formation by core fragmentation in the first stages of the stellar system formation. These mechanisms have not yet been proven to occur for brown dwarfs, making \tstar an excellent system to study in detail.

\begin{acknowledgements}
Based on observations collected at the European Organisation for Astronomical Research in the Southern Hemisphere under ESO programme 105.20ND. 
This work is partly financed by the Spanish Ministry of Economics and Competitiveness through project PGC2018-098153-B-C31. 
R.\,L. has received funding from the European Union’s Horizon 2020 research and innovation program under the Marie Skłodowska-Curie grant agreement No.~713673 and financial support through the “la Caixa” INPhINIT Fellowship Grant LCF/BQ/IN17/11620033 for Doctoral studies at Spanish Research Centers of Excellence from “la Caixa” Banking Foundation, Barcelona, Spain. 
M.\,R.\,Z.\,O. acknowledges financial support from projects AYA2016-79425-C3-2-P and PID2019-109522GB-C51 funded by the Spanish Ministry of Science, Innovation and Universities. 
This publication makes use of VOSA, developed under the Spanish Virtual Observatory project supported by the Spanish MINECO through grant AyA2017-84089. VOSA has been partially updated by using funding from the European Union's Horizon 2020 Research and Innovation Programme, under Grant Agreement nº 776403 (EXOPLANETS-A). 
The Pan-STARRS1 Surveys (PS1) and the PS1 public science archive have been made possible through contributions by the Institute for Astronomy, the University of Hawaii, the Pan-STARRS Project Office, the Max-Planck Society and its participating institutes, the Max Planck Institute for Astronomy, Heidelberg and the Max Planck Institute for Extraterrestrial Physics, Garching, The Johns Hopkins University, Durham University, the University of Edinburgh, the Queen's University Belfast, the Harvard-Smithsonian Center for Astrophysics, the Las Cumbres Observatory Global Telescope Network Incorporated, the National Central University of Taiwan, the Space Telescope Science Institute, the National Aeronautics and Space Administration under Grant No. NNX08AR22G issued through the Planetary Science Division of the NASA Science Mission Directorate, the National Science Foundation Grant No. AST-1238877, the University of Maryland, Eotvos Lorand University (ELTE), the Los Alamos National Laboratory, and the Gordon and Betty Moore Foundation.
This work is partly supported by JSPS KAKENHI Grant Numbers JP18H01265 and JP18H05439, and JST PRESTO Grant Number JPMJPR1775. 
H.\,M.\,T. acknowledges financial suppport from FCT - Fundação para a Ciência e a Tecnologia through national funds and  FEDER through COMPETE2020 - Programa Operacional Competitividade e Internacionalização under these grants: UID/FIS/04434/2019, UIDB/04434/2020; UIDP/04434/2020, PTDC/FIS-AST/28953/2017, and POCI-01-0145-FEDER-028953. A.\,J.\,M.\ is supported by the Swedish National Space Agency (career grant 120/19C).

\end{acknowledgements}

\bibliographystyle{aa.bst} 
\bibliography{biblio.bib}

\end{document}